\documentstyle[multicol,aps,psfig]{revtex}
\begin{document}

\draft

\preprint{\vbox{\hbox{DRAFT of U. of Iowa preprint 2000-2503}}}

\title{Small Numerators Canceling Small Denominators: Is Dyson's 
Hierarchical Model Solvable? }

\author{Y. Meurice \\ 
{\it Department of Physics and Astronomy, The University of Iowa, 
Iowa City, Iowa 52242, USA}}

\maketitle

\begin{abstract}

We present an  analytical method to solve 
Dyson's hierarchical model, involving 
the scaling variables 
near the high-temperature fixed point. The procedure
seems plagued by the presence of small denominators
as in perturbative expansions near integrable systems in Hamiltonian mechanics.
However, in all cases considered,
a zero denominator always comes with a zero numerator.
We conjecture that these cancellations occur in general,
suggesting that the model has remarkable features reminiscent 
of the integrable systems.
\end{abstract}
\pacs{PACS: 05.50.+q, 11.10.Hi, 64.60.Ak, 75.40.Cx}
\begin{multicols}{2}\global\columnwidth20.5pc
\multicolsep=8pt plus 4pt minus 3pt
 
In many physical problems involving nonlinear flows, a common
strategy consists in constructing a system of coordinates where 
the flow becomes linear. The action-angle variables in Hamiltonian mechanics 
provide well-know examples of such a procedure. Whenever the angle
variables can be constructed, they evolve linearly with time and 
expressing the original variables in terms of the new ones 
solves completely the original problem.
The problems for which 
well-defined angle-action variables can be constructed 
(e.g. Kepler's problem or the free rigid body) are 
very distinguished and
called integrable systems. 
A large number of numerical experiment has lead us to 
believe that in a generic way, 
small perturbations destroy integrability. 
This point of view was was first inferred
by H. Poincar\'e who pointed  out the existence of small denominators 
in the canonical transformation designed to eliminate the angle
dependence of a perturbed Hamiltonian. 

In this letter, we discuss the question of small denominators for
renormalization group (RG) flows. 
The variables which play the role of angle variables are the
scaling variables introduced by Wegner \cite{wegner72}. Near a fixed point, 
the RG flows can be linearized. The problem of expressing the physical 
quantities in terms of 
variables which transform as in the linear approximation 
when the non-linear 
terms are taken into account, is analogous to removing the 
angle dependence of a perturbed Hamiltonian. If the task
can be carried through, one obtains analytical expressions 
for the RG flows. 
However, as we will show, small denominators appear.
Does this mean that, as in Hamiltonian mechanics, 
in generic situations the construction 
will fail?

Surprisingly, we found in a numerical calculation performed with
Dyson's
hierarchical model \cite{dyson69,baker72},
that zero denominators were systematically 
canceled by zero numerators. These remarkable cancellations
suggest that either the model considered
is as distinguished
as the integrable systems of classical mechanics or 
that there exists a general mechanism that allows us to 
circumvent the small denominator problem for RG flows. 
The example of the two-dimensional Ising model shows the 
importance of having a non-trivial model which can
solved in closed form. The results presented below indicate 
that Dyson's hierarchical can be solved analytically.

During the last decades, the RG method has been successfully applied
to many important problems in field theory and statistical mechanics:
the critical behavior of ferromagnets and superconductors,
the confinement of quarks or the generation of mass for the W and Z 
bosons.
However, its practical implementation 
is still a formidable enterprise.
Typically, expansions are often available near fixed points,
but not to orders large enough to allow one to extrapolate between 
fixed points. Unfortunately, the calculation of physical quantities 
(e. g. the magnetic susceptibility) beyond an 
order of magnitude estimation, requires a calculation of the 
flows in crossover regions. 
One has then to rely to Monte Carlo  simulations to achieve this goal.
One also needs to select a small set of interactions which closes
reasonably well under RG transformation near both fixed points.
Interesting examples of such lattice Monte Carlo calculations are given in
Refs. \cite{gonzalez87} for scalar theory
and \cite{taro00} for gauge theories. 

In order to get analytical results, further approximations are
needed. One possibility consists in using hierarchical 
approximations such as the one derived by Wilson in Ref. \cite{wilson71b}
and resulted into the ``approximate recursion formula''. 
In this approximation, only the local interactions get renormalized
and the flow can be calculated from a simple integral formula. 
Retracing Wilson's construction from the beginning, one can restore 
the other renormalizations perturbatively. In order to perform this task,
one would like to have a closed form solution in the hierarchical 
approximation. In order to achieve this goal, we have proposed \cite{scaling} 
to use the Fourier representation of the integral formula and to 
construct the scaling variables in this basis. In this process, we 
identified the existence of small denominators possibly 
ruining the whole approach. This 
problem can be avoided in special circumstances, for instance 
for flows starting exactly along the unstable direction
of a non-trivial fixed point. With this restriction, we found \cite{scaling}
expansions with overlapping domains of convergence in the crossover region.
 
In the following, we discuss the problem of small denominators 
in the construction of the scaling variables near the 
high-temperature (HT) fixed point  of Dyson's hierarchical model.
The treatment of the this model is
mathematically similar to the one of the approximate recursion formula.
However there exists a large literature on
Dyson's model \cite{baker77,collet78,bleher75}, the non-trivial 
fixed point is known very precisely \cite{koch95} and the HT 
expansion studied to a very large order \cite{osc1}. 

For a description of this model as a spin model, we refer to 
Ref. \cite{finite,hyper}, while the details of 
the derivation of the RG flows as expressed
below can be found in \cite{scaling}. To make a long story short, all 
the information regarding the local interactions 
after $n$ RG transformations is encoded in a function
$R_n(k)=1+a_{n,1}k^2+a_{n,2}k^4+\dots$. The logarithm of this function 
generates the connected zero-momentum Green's functions at finite volume.
The recursion formula reads 
\begin{equation}
R_{n+1}(k) = C_{n+1} \exp \left[ -{1\over 2} 
{{\partial ^2} \over 
{\partial k ^2}} \right]\left[R_{n} \left({\sqrt{c}k\over 2} \right) \right]^2 \ . 
\label{eq:rec}\end{equation}
We fix the normalization constant $C_{n}$ so that $R_{n}(0) = 1$.
We use the parametrization $c=2^{1-2/D}$ which implies that a free
massless field scales in the same way as in a 
usual $D$-dimensional theory. 
In this formulation, the temperature dependence has been absorbed in the 
initial $R_0(k)$.
For an Ising measure, $R_{0}(k) = \cos(\sqrt{\beta}k)$, while in general,
we have to numerically integrate the Fourier transform of 
the local measure to determine the coefficients of
$R_{0}(k)$ expanded in terms of $k$. In general, $a_{n,l}$ is of order
$\beta^l$ in the HT expansion.

In the HT phase, polynomial truncations of order $l_{max}$
in $k^2$ provide rapidly converging approximations \cite{finite}.
The RG flows can be expressed in terms of the quadratic map
\begin{equation}
a_{n+1, l} = \frac{u_{n,l}}{u_{n,0}} \ ,
\label{eq:aofu}
\end{equation}
with
\begin{equation}
u_{n,\sigma} = \Gamma_{\sigma}^{ \mu \nu} a_{n,\mu} a_{n,\nu} \ ,
\end{equation}
and
\begin{equation}
\Gamma_{\sigma}^{ \mu \nu}
= (c/4)^{\mu+\nu}\ 
\frac{(-1/2)^{\mu + \nu - \sigma}(2(\mu+\nu))!}{ 
(\mu+\nu-\sigma)!(2 \sigma)!}  \ ,
\label{eq:struct}
\end{equation}
for $\mu+\nu \geq\sigma$ and zero otherwise.
We use ``relativistic'' notations. 
Repeated indices mean summation.
The greek indices $\mu$ and $\nu$
go from $0$ to $l_{max}$, while latin indices $i$, $j$ go from 1 to   
 $l_{max}$. 

The diagonalization of the linear RG transformation near the HT fixed 
point is quite simple because it is of the upper triangular form.
From Eq. (\ref{eq:struct}), 
one finds the spectrum 
\begin{equation}
\lambda_{(r)}=2(c/4)^{r} \ .
\label{eq:hteigenv}
\end{equation}
in agreement with Ref. \cite{collet78}.
Using the matrix of right eigenvectors, ${\cal M}_{l}^{i} \psi^r_i = \lambda_{(r)} \psi^r_l$, 
we introduce new coordinates such that 
$a_{n,l} = \psi^r_l h_{n,r}$. This diagonalizes the linear RG transformation.
Note that the form of the eigenvectors guarantees that 
$h_{n,l}$ is also of order $\beta^l$. 
More details are given Ref. \cite{scaling}, where all the quantities 
related to the HT fixed point are dressed with a ``tilde'' omitted here .
In summary, the RG flows in the new coordinates can be written as
\begin{equation}
h_{n+1,l} = \frac{\lambda_{(l)} h_{n,l}
 + \Delta_{l}^{ p q} h_{n,p} h_{n,q} }
{1 + \Lambda^{p} h_{n,p} 
+ \Delta_{0}^{ p q} h_{n,p} h_{n,q}}\ ,
\label{eq:hrules}
\end{equation}
with coefficients calculable from Eq. (\ref{eq:struct}).

We now express the $h_{n,l}$ in terms of
the scaling variables $y_{n,1},\ldots,
y_{n,l_{max}}$ which
transform under a RG transformation as $y_{n+1,i} = \lambda_{i} y_{n,i}$.
If we can construct functions $h_l$ and $y_l$ such that 
$h_{n,l}=h_l({\mathbf{y}}_{n})$ and  $y_{n,l}=y_l({\mathbf{h}}_{n})$,
then we get a complete analytical expression of $h_{n,l}$ (which contains
all the thermodynamical quantities)
in terms 
of $h_{0,l}$ (which depends on the initial energy density):
\begin{equation}
h_{n,l}=h_l(\lambda_1^ny_1({\mathbf h}_0),\lambda_2^ny_2({\mathbf h}_0),
\dots)\ .
\label{eq:solve}
\end{equation}
The feasibility of this approach is demonstrated for a 
one-dimensional example in 
Ref. \cite{dual}.

We now discuss the construction of the $h_l$. We use the expansion
\begin{equation}
h_{l} = y_l+\sum_{i_{1},i_{2},\ldots} s_{l,i_{1} i_{2} \ldots} y_{1}^{i_{1}}
y_{2}^{i_{2}} \ldots \ ,
\end{equation}
where the sums over the $i$'s run from $0$ to infinity in each variable
with at least two non-zero indices.
In the following, we use the notation ${\mathbf{i}}$ for $(i_1,i_2,\dots)$.
Plugging the expansion into Eq. (\ref{eq:hrules}), 
and requiring that the one step advance 
is obtained by rescaling the scaling variables by their associated
eigenvalue, we obtain
\begin{equation}
s_{l, \mathbf{i}} = \frac{N_{l,\mathbf{i}}}
{\left(\prod_{m} \lambda_{(m)}^{i_{m}} - \lambda_{r}\right)} \ .
\label{eq:denom}
\end{equation}
with
\begin{eqnarray}
\label{eq:num}
N_{l, \mathbf{i}} =& \sum_{\mathbf{j}+\mathbf{k} = \mathbf{i}}
 ( \Delta_{l}^{ p q} s_{p,\mathbf{j}}
 s_{q,\mathbf{k}} - s_{l,\mathbf{j}} \prod_{m} \lambda_{(m)}^{j_{m}}
\Lambda^{p} s_{p, \mathbf{k}} )\\ \nonumber
&-\sum_{\mathbf{j}
+\mathbf{k}+\mathbf{r}=\mathbf{i}}
 s_{l, \mathbf{j}} \prod_{m} \lambda_{(m)}^{j_{m}} \Delta_{0}^{ p q}
 s_{p, \mathbf{k}} s_{q, \mathbf{r}} \ .
\end{eqnarray}
For a given set of indices $\mathbf{i}$, we
introduce the notation 
\begin{equation}
{\cal{I}}_q ({\mathbf i})=\sum_m i_m m^q \ .
\label{eq:indices}
\end{equation}
One sees that ${\cal{I}}_0$ is the degree of the associated 
 monomial and ${\cal{I}}_1$
its order in the HT expansion (since $y_l$ is also of order $\beta ^l$). 
Given that all the indices are positive and that at least one index 
is not zero,
one can see that if ${\mathbf{j}}+{\mathbf{k}}={\mathbf{i}}$ then
${\cal I}_q({\mathbf{j}})<{\cal I}_q({\mathbf{i}})$ and 
${\cal I}_q({\mathbf{k}})<{\cal I}_q({\mathbf{i}})$. Consequently,
Eq. (\ref{eq:num}) provides a solution order by order in ${\cal{I}}_0$ or
in ${\cal{I}}_1$ (since the r.h.s is always of lower order) 
assuming that the small denominator problem can be avoided.

Using the parametrization $c=2^{1-2/D}$, 
the zero denominators in Eq. (\ref{eq:denom}) appear when
\begin{equation}
D-l(D+2)=D{\cal{I}}_0 -(D+2){\cal{I}}_1 \ .
\end{equation}
If $D$ and $D+2$ have no common factors, this equation has non-trivial 
solutions for indices such that ${\cal{I}}_0=(2+D)q+1$, for $q$ a strictly
positive integer and $l$ such
that ${\cal{I}}_1=Dq+l$. If $D$ and $D+2$ have common factors, we can proceed
in the same way but with these common factors removed from both $D$ and $D+2$.

We have investigated numerically the well-studied case $D=3$
with a polynomial truncation $l_{max}=25$. 
For practical reasons, we have limited our study the case where 
$h_l$ depends only on 
the two leading variables $y_1$ and $y_2$. This restriction is self-consistent
since if we start for instance with $y_3=0$, the multiplicative
renormalization of the scaling variables implies that this condition 
stays valid after $n$ iterations. Since $i_3=i_4=\dots=0$,
we will use notations such as $s_{l,i_1,i_2}$ or $h_l(y_1,y_2)$.
In addition, we have concentrated our attention to the large order
behavior in the leading variable $y_1$ and calculated the
first order corrections in the subleading variable $y_2$
If $i_2=0$, we have the following sets of small denominators:
$i_1=5q+1$ and $l=2q+1$. If $i_2=1$, we have $i_1=5q$ and $l=2q+2$.
We have calculated the numerators corresponding to these zero 
denominators for $q=1$, 2, 3 and 4. 

We have checked our calculation of the $h_l$
with two different methods. First we have
considered a configuration
$h_{n,l}=h_l(y_1,y_2)$
for particular values of $y_1$ and $y_2$ and then calculated 
the one step backward configuration
$h_{n-1,l}=h_l(\lambda_1^{-1}y_1,\lambda_2^{-1}y_2)$.
We then calculated the one step forward using these $h_{n-1,l}$
and the exact formula Eq. (\ref{eq:hrules}). 
Comparison between the two $h_{n,l}$ 
for values of $y_1$ and $y_2$ varying between 0.1 and 0.001 and
for various truncations in the power of $y_1$ and $y_2$ considered,
showed errors scaling like the powers neglected with coefficients
compatible with the order of magnitude of the coefficients involved in 
the expansion. Second, we used the expression $y_1(h_l)$ calculated in
Ref. \cite{scaling} up to order 11 in $\beta$
and checked $y_1(h_l(y_1,y_2))=y_1$ with errors 
ranging between
$10^{-14}$ and $10^{-16}$ for the various coefficients of the 
higher order terms. 

In Fig. 1 we show the absolute value of $N_{3,j,0}$. The calculations
have been performed with different arithmetic precisions.
We have used $Mathematica$ 4.0
with $\$MaxPrecision$ and $\$ MinPrecision$ 
both set to the same value $pr$.
We have considered the cases $pr$=20, 30 and 40. 
The graphs shows $N_{3,6,0}$ is 
more than twenty orders of magnitudes smaller than its peers when
$pr=20$ and then drops by ten orders of magnitude each time the precision 
is increased by 10, while all the other values stay stable. This 
is a strong evidence for $N_{3,6,0}=0$
\begin{figure}
\centerline{\psfig{figure=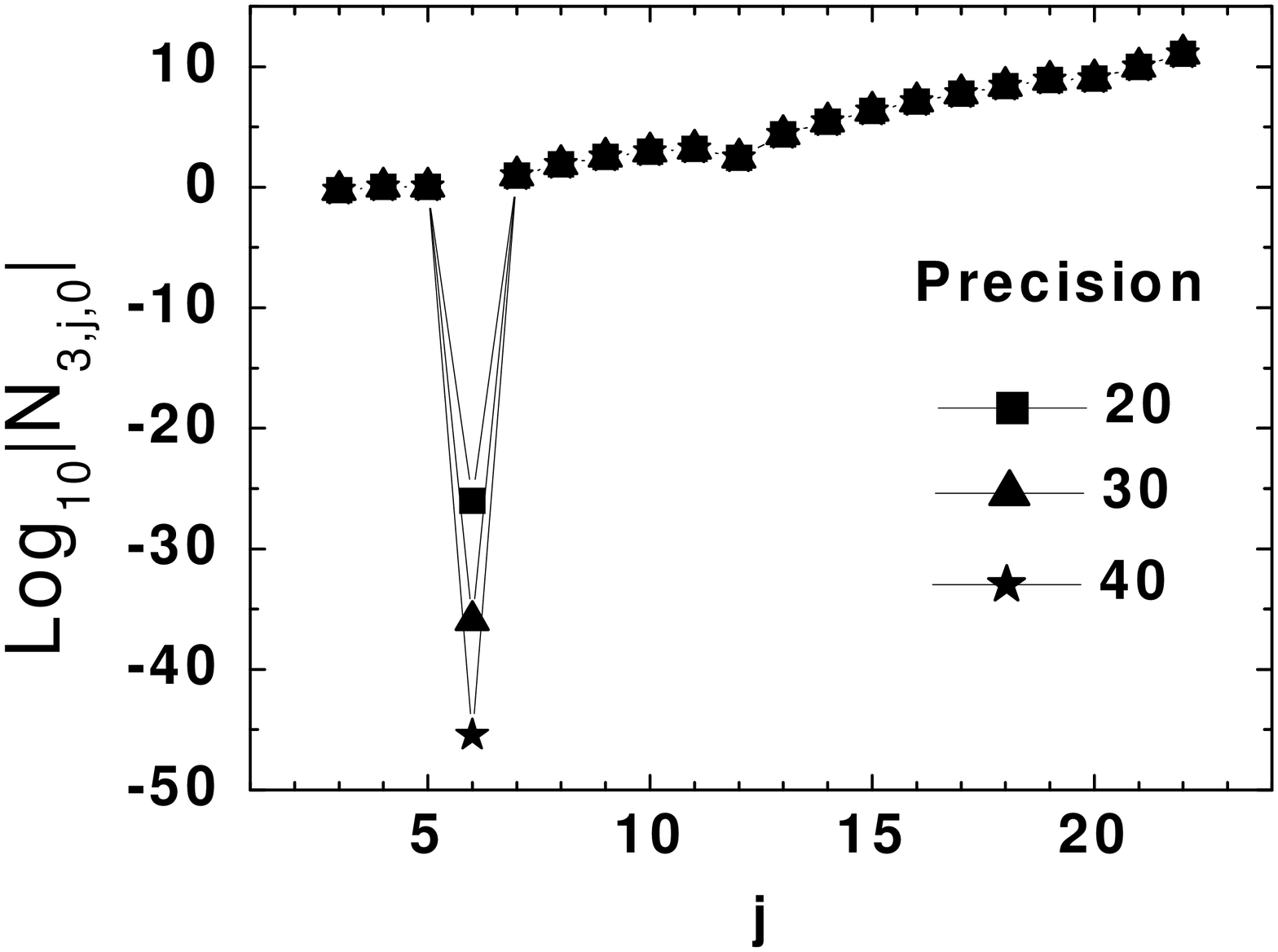,width=3in}}
\caption{$Log_{10}(|N_{3,j,0}|)$ versus $j$.}
\label{fig:36}
\end{figure} 
In Figs. \ref{fig:all0} and \ref{fig:all2}, we show
$N_{l,i_1,0}$ 
and  $N_{l,i_1,1}$ calculated with
$pr$=20.
One sees that $N_{2q+1,5q+1,0}$ and $N_{2q+2,5q,1}$ are more than
20 orders of magnitude smaller than the naive interpolation.
We have checked in each of these cases that
the small value drops by ten order of magnitudes each time the 
precision is increased just as in Fig. \ref{fig:36}. 
\begin{figure}
\centerline{\psfig{figure=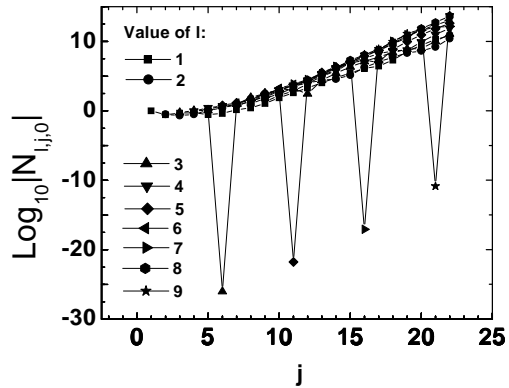,width=3in}}
\caption{$Log_{10}(|N_{l,j,0}|)$ versus $j$ for $l=1,\dots 9$.}
\label{fig:all0}
\end{figure} 
\begin{figure}
\centerline{\psfig{figure=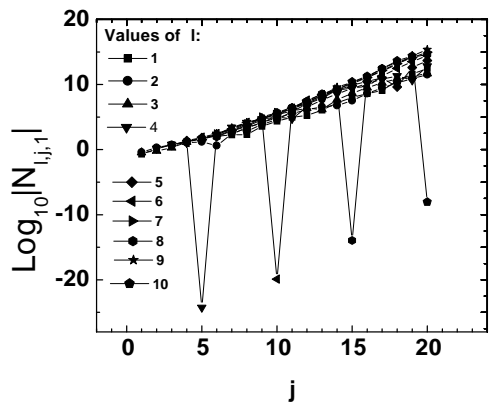,width=3in}}
\caption{$Log_{10}(|N_{l,j,1}|)$ versus $j$ for $l=1,\dots 10$.}
\label{fig:all2}
\end{figure} 

In summary, for
each zero denominator considered, we found a
zero numerator. 
We thus conjecture that all numerators corresponding to zero denominators 
are zero. 
If this conjecture is correct, the $s_{l,{\bf i }}$ corresponding to 
non-zero denominators are calculable from Eq. (\ref{eq:num}) 
and those corresponding to 
zero denominators are
undetermined. The calculations 
performed above have been done with these coefficients treated as undetermined.
The figures have been drawn with these coefficients set to zero.
This choice is not essential, we have considered other choices where 
the undetermined coefficients have been set to values  of 
the same order as the coefficients above and below (in $i_1$) and reached 
identical conclusions. This is a non-trivial statement. For instance, 
the undetermined coefficient
$s_{3,6,0}$ appears explicitly in the numerator of the 
equation for $s_{5,11,0}$, however it is multiplied by a very small 
number which drops when $pr$ is increased. 

We have checked that the individual terms in the 
zero numerators were 
not zero. 
The fact that we were able to obtain very precise cancellations 
without any fine-tuning
suggests the existence of 
closed form formulas or of a symmetry forbidding these terms. More
generally, the equally spaced spectrum (on a log scale), the number of terms
at a given level $m$ (the order in HT expansion) 
equal to the number of partitions of $m$
\cite{scaling}
and the existence of ``nested'' ambiguities are somehow reminiscent
of string theory. A possible starting point to discover these 
hypothetical symmetries would be to exploit the fact that since
for instance $y_3$ and $y_1^6$ transform the same way under 
a RG transformation, there exist ambiguities in the 
construction
of $h_l$ in terms of the scaling variables.

We  should say a few words about the small denominators
near other fixed points. The spectrum of the linearized RG transformation 
near the non-trivial fixed point \cite{koch95} 
can be calculated numerically \cite{gam3rapid}. 
Investigating the small denominators 
of the form $\lambda_l^k\simeq \lambda_m^r$ for the first twenty eigenvalues
and for powers not larger than 100. The best solution we found was 
$\lambda_2^9\simeq\lambda_4$ with two parts in a thousand. 
A more detailed study is necessary to decide if the rate of decay 
of the coefficients is sufficient to take care of the smaller denominator
which will appear at larger orders.
On the other hand, the spectrum at the Gaussian fixed point \cite{collet78}
is $\lambda_j=2c^{-j}$. There are many zero denominators in integer
dimensions,
 e.g., $\lambda_1=\lambda_2^2$ for $D=3$. 
If the numerators are not zero, one
can ``repair'' \cite{wegner72} the situation by considering $n$-dependent
coefficients. This excludes a solution of the form of 
Eq. (\ref{eq:solve}) but it 
generates logarithmic corrections which are 
necessary. These can even be observed in the large order
of the HT temperature expansion in Ref. \cite{ht4}.

In conclusion, we have found remarkable cancellations of small
denominators by small numerators. If these cancellations occur 
in general, it is possible to calculate analytically the 
thermodynamical quantities of Dyson's hierarchical model in the HT phase
using Eq. (\ref{eq:num}). 
Our results  suggest that this model 
has features analogous to the integrable systems in Hamiltonian
mechanics. As such, the model would stand out as a first 
approximation to be used in situations where the conventional
perturbative expansions are not reliable.

This research was supported in part by the Department of Energy
under Contract No. FG02-91ER40664.
Y. M. thanks the 
Aspen Center for Physics for its hospitality in Summer 2000 while 
this work was in progress and for a conversation there with L. Kadanoff.


\begin{thebibliography}{10}
\bibitem{wegner72}
F. Wegner, Phys. Rev. B {\bf 3},  4529  (1972).

\bibitem{dyson69}
F. Dyson, Comm.\ Math.\ Phys.\ {\bf 12},  91  (1969).

\bibitem{baker72}
G. Baker, Phys.\ Rev.\ B {\bf 5},  2622  (1972).

\bibitem{gonzalez87}
A. Gonzalez-Arroyo and M. Okawa, Phys. Rev. D {\bf 35},  672  (1987).

\bibitem{taro00}
P. de~Forcrand~et al., Nucl. Phys. B {\bf 577},  263  (2000).

\bibitem{wilson71b}
K. Wilson, Phys. Rev. B. {\bf 4},  3185  (1971).

\bibitem{scaling}
Y. Meurice and S. Niermann, u. of Iowa Preprint, hep-lat/00077037.

\bibitem{baker77}
G. Baker and G. Golner, Phys. Rev. B {\bf 16},  2080  (1977).

\bibitem{collet78}
P. Collet and J. Eckmann, {\em A Renormalization Group Analysis of the
  Hierarchical Model in Statistical Mechanics} (Springer-Verlag, Berlin, 1978).

\bibitem{bleher75}
P. Bleher and Y. Sinai, Comm. Math. Phys. {\bf 45},  247  (1975).

\bibitem{koch95}
H. Koch and P. Wittwer, Math. Phys. Electr. Jour. {\bf 1},  Paper 6  (1995).

\bibitem{osc1}
Y. Meurice, G. Ordaz, and V.~G.~J. Rodgers, Phys.\ Rev.\ Lett. {\bf 75},  4555
  (1995).

\bibitem{finite}
J. Godina, Y. Meurice, M. Oktay, and S. Niermann, Phys. Rev. D {\bf 57},  6326
  (1998).

\bibitem{hyper}
J.~J. Godina, Y. Meurice, and M. Oktay, Phys. Rev. D {\bf 61},  114509  (2000).

\bibitem{dual}
Y. Meurice and S. Niermann, Phys. Rev. E {\bf 60},  2612  (1999).

\bibitem{gam3rapid}
J. Godina, Y. Meurice, and M. Oktay, Phys. Rev. D {\bf 57},  R6581  (1998).

\bibitem{ht4}
J.~J. Godina, Y. Meurice, and S. Niermann, Nucl. Phys. B {\bf 519},  737
  (1998).

\end{thebibliography}

\end{multicols}
\end{document}